# Electronic Transport and Magnetic Properties of Layer-Antiferromagnet $CuCrS_2$

## Abstract

We have performed a detailed study of the electrical and thermal conductivities and thermoelectric power behavior of an antiferromagnetic layer compound of chromium, $CuCrS_2$, from 15K to 300K. Unlike previous studies, we find non-insulating properties and a sensitive dependence on the preparation methods, the microstructure and the flaky texture formed in polycrystalline samples after extended sintering at high temperatures. Flakes are found to be metallic with strong localization effects in the conductivity on cooling to low temperatures. The antiferromagnetic transition temperature $T_N$ (=40K) remains essentially unaffected. The Seebeck coefficient is found to be in the range 150-450 µV/K, which is exceptionally large and becomes temperature independent at high temperatures, even for the specimen with a low resistivity value of 5-200 mΩ-cm. We find the thermal conductivity $\kappa$ to be low, viz. 5-30mW/K-cm. This can be attributed mostly to the dominance of lattice conduction over the electronic. The value of $\kappa$ is further reduced by the disorder in Cu-occupancy in the quenched phase. We also observe an unusually strong dip in $\kappa$ at $T_N$ which is probably due to strong magnetocrystalline coupling in these compounds. Finally we discuss the properties of $CuCrS_2$ as a heavily doped Kondo-like insulator in its paramagnetic phase. The combination of the electronic properties observed in $CuCrS_2$ makes it a potential candidate for various thermoelectric applications.

## Introduction

The ternary compounds of chromium, the cubic thiospinels $ACr_2X_4$ (A = Cu, Zn, Cd or Hg and X =S, Se or Te) and structurally related hexagonal layered $ÁCrX_2$ (Á = Li, Na, K, Cu or Ag) of the present study, have been extensively investigated for their magnetic properties[1-3]. In these compounds the magnetic interactions have important effect on the electronic transport. The Cr-thiospinels are well known magnetic semiconductors. However, among these the compounds containing Cu, i.e., $CuCr_2S_4$ ($Se_4$, $Te_4$) are metallic with high values of ferromagnetic ordering temperature $T_C$ of 420K, 460K and 365K respectively. The metallic nature and the ferromagnetic behavior of these compounds have been explained in terms of intermediate valency of Cr-atoms[1,4,5] which leads to a double exchange mechanism, as was proposed by Zener.

The hexagonal layered compounds of Cr including $CuCrS_2$ ($Se_2$) are antiferromagnetic. The ordering temperatures $T_N$ in these compounds are low (20K - 55K) despite the strong magnetic interactions, as can be deduced from the high value of Curie-Weiss temperature $\theta_{CW}$ which varies from +250K for $KCrSe_2$[6] to –250K for $LiCrS_2$[7] in their paramagnetic phases. $\theta_{CW}$ is related to the Cr-Cr distances within the hexagonal layer, and its strong variation in different compounds has been attributed to the competing interactions between the direct antiferromagnetic and ferromagnetic super–exchange through anion. The low value of $T_N$ is due to weak antiferromagnetic interactions between successive Cr-layers through the intervening Cu-layer. This interpretation assumes a localized and insulating nature of electrons of $Cr^{3+}$- ions that have spin only moment of $3.9\mu_B$ for $d^3$ configuration. Our present study, however, shows non-insulating properties and does not support purely localized picture for the exchange interactions in these compounds. Moreover, the neutron scattering studies have revealed that the ordered Cr-moments are about 20% less than expected value of $3.9\mu_B$ and the direction of moment is modulated with the modulating vectors depending on the Cr–Cr distances in the different compounds. The successive planes in all of them are coupled antiferromagnetically[7, 8].

The electronic transport properties of $CuCrS_2$ and $CuCrSe_2$ have not been investigated in detail. A preliminary conductivity measurement on the crystals of $CuCrS_2$ by Le Nagard et al [9] indicated a semiconducting behavior. They found unusually low activation energy $E_{act}$ of around 4meV in the temperature range 60K - 300K and a large conductivity at lower temperatures, thereby indicating non-insulating properties. Another study suggested that $CuCrS_2$ is a non-stoichiometric semiconductor and exhibits changeover from p-type to n-type when prepared under a reduced sulfur-vapor pressure and at higher temperatures[10]. This study was performed at temperatures above 600$^0$C, so their conclusions are not applicable for the low temperature transport properties. Moreover, Cu(Ag)CrS$_2$ also shows an order-disorder transition of Cu-atoms around 400$^0$C that may affect the electronic transport properties at high temperatures[7]. Apart from these studies, Cu(Ag)CrS$_2$ has also been characterized as a mixed ion-electron conductor. A large diffusion ionic conductivity of Cu$^{+1}$- ions on the tetrahedral sites was measured above room temperature[11]. As far as the electronic conductivity is concerned, one study found very large resistivity ~ 10$^4$ Ω-cm and the other an immeasurable high resistivity in CuCrS$_2$ at room temperature [12].

We have measured the electronic transport properties of CuCrS$_2$ prepared by different methods. The electronic transport properties of CuCrS$_2$ with temperature and crystalline quality are found to be qualitatively different from the previous studies mentioned above. We also report, for the first time the results of thermoelectric power and thermal conductivity measurements on these compounds. The combination of our results shows that these compounds are potential candidates for various thermoelectric applications.

## Preparation and Characterization

The compounds were prepared directly from the reaction of elements Cu(99.9%), Cr(99.9%) and S(99.9%) and also from the diffusive reaction of homogenized and compressed mixture of their binary compounds CuS(99+%) and Cr$_2$S$_3$(99%) in evacuated and sealed quartz tubes. We have used CuS instead of more stable Cu$_2$S used in previous studies [9,12] and have carried out a very slow diffusive reaction of compressed mixture of binaries at a lower temperature of about 650$^0$C for five days. This procedure was adopted to avoid the formation of ferromagnetic phase CuCr$_2$S$_4$. The excess sulfur used in this method was separated and deposited at the cold end of the tube. The layered compounds Cu$_x$CrS$_2$: x=0.9, 0.8, free from ferromagnetic spinel inclusions, could also be prepared by this method by varying the amount of CuS in the starting mixture. The regrinding, pelletizing under 5-ton pressure and annealing at 850$^0$C were necessary to get strain free homogeneous final products. Some pellets were quenched from 850$^0$C in air (sample 1 and 2) and others slowly cooled (sample 3) to room temperature to obtain the quenched and annealed phases of the respective compositions. The extended period of sintering at high temperature gave highly textured pellets (samples 1, 2 and 3) with the growth of crystal-flakes parallel to the surface of the pellets, as can easily be seen from the x-ray diffraction pattern (Figure 1). On the other hand, the texturing was absent in the phases which were synthesized by the diffusive reaction of CuS and $Cr_2S_3$ (samples 4 and 5).

The single crystal flakes of sizes ~5x5x0.05 $mm^3$ were obtained along the lower temperature end of a long tube that was kept at the temperature of 700-800$^0$C for 3-4days. We have detected approximately 3-4% of ferromagnetic impurity phase in the flakes and powders, when they were prepared at lower temperatures and by using excess amount of sulfur. The magnetic impurities were significantly reduced to less than 1% in our flakes after re-annealing them at 850$^0$C for 4-5 hours in a running vacuum of $10^{-2}$ mm of Hg. The phases obtained at high temperature were found to be cation rich. Typical atomic ratio of Cu: Cr: S was found to be equal to 1.08:1.07:2 from the EDAX analysis. We could not ascertain the off-stoichiometry of our flakes because of small inclusions of the magnetic phase in them.

**X-ray diffraction:** The compounds prepared by different methods, quenched and annealed specimen as well as $Cu_xCrS_2$: x = 0.8, 0.9 , gave sharp peaks in x-ray diffraction pattern without any significant change in the cell-parameters from the previously reported values by Le Nagard et al for the crystals [9]. We also support their conclusion that a large disorder in the Cu-occupancy in this compound cannot be quenched below the ordering transition at 400$^0$C, since no discernible changes in X-ray peaks were noticed by rapid cooling through the transition.

*Platy texture:* In figure 1, we show the X-ray pattern that was recorded by using the reflections directly from the surface of the pellet that was sintered over long period of 5 days at 850$^0$C. We compare this pattern with the pattern shown in figure 1 expected for the randomly oriented crystallites in its powder [9]. The increased intensity of all the (00l) reflections shows the platy texture of the surface of pellet. We will see below that this texturing has significant effects on the transport properties of the pellets.

**Magnetic properties**

The magnetic susceptibility $\chi$ of $CuCrS_2$ from 2K to 300K at 500 Oe field, and the magnetization (in the inset of figure 2) up to 14T field at 2K are plotted in figure 2. $\chi$ follows Curie-Weiss dependence, $\chi = C/(T+\theta)$ ; $C$ = 1.9 (emu/mole) and $\theta$ = 110K, above 200K shown by a continuous line in figure 2. The *C* corresponds to an effective moment of 3.9 $\mu_B$ / Cr as expected for total spin 3/2 for its $d^3$ configuration and with the quenched orbital contribution. The susceptibility $\chi$ starts deviating from the Curie–Weiss dependence below 200K on cooling towards the magnetic transition $T_N$ = 40.5K. The magnetization in the antiferromagnetic phase at 2K up to 14T field is plotted in the inset of figure 2; it shows an upward curvature at high fields. This $M(H)$ dependence confirms the helicoidal order of Cr moments, earlier deduced from the neutron scattering measurements [8]. Significantly, the neutron scattering also gave the moments on the Cr-atoms that were about 20% smaller in the ordered phase from the expected value for spin 3/2 in its paramagnetic phase. There is no explanation available for the reduction of the Cr-moments at low temperatures. In our opinion the observation of fractional moments of Cr-atoms in $CuCrS_2$ is quite significant and may be the consequence of the non-localized nature of their electrons, resulting from the strong hybridization of $3d$-orbital of Cr with the $sp$-orbital of surrounding sulfur atoms.

## Transport properties

*Experimental details:* The electrical conductivity measurements on sintered pellets and crystal flakes between 15K and 300K were performed by four probe method using silver paste for the electrical contacts. The thermopower of the pellets and the crystal flakes was studied by measuring Seebeck coefficient $S$ (= $\Delta V/\Delta T$) with respect to copper from 15K to 300K using a differential method of measurement. In this method, at a stabilized temperature a small temperature gradient is generated across the sample length and the thermoelectric voltage is recorded using copper leads. The spurious and offset voltages of the measuring circuit were eliminated by reversing the temperature gradient and averaging the recorded voltages. The apparatus was tested for the accuracy by measuring $S$ on a thin piece of pure Lead with respect to copper versus temperature[13]. The thermal-conductivity $\kappa$ at different temperatures was measured by passing a known amount of heat current through the length of a rectangular piece of pellet and recording the temperature difference by a differential thermocouple (Au (0.05%Fe) / Chromel) after steady state was reached. This method was tested by measuring $\kappa$ of a rectangular piece of sintered Alumina–substrate from 15K to 300K (shown in the inset of figure 5) and comparing it to the expected values[14]. The accuracy of thermopower and thermal conductivity values for our compounds are about 5%.

*Electrical conductivity:* In figure 3, we show temperature dependence of resistivity of different samples. The resistivity of the samples prepared directly from the elements (1, 2 and 3) and also from the inter diffusion of binary compounds (4 and 5) is plotted as log ($\rho$) in panel 'a' and the in-plane resistance of some of the crystal flakes is shown on a linear scale in panel 'b'. The final heat treatment in case of samples 1, 2 and 3 was sintering at 850-900$^0$C for 5 days, followed by quenching (1 and 2) and slow cooling (3). The pellets of all of them showed varying degree of flaky-texturing as mentioned above. Sample 3 was prepared using 10% excess sulfur, and it contained small amount (<2%) of ferromagnetic inclusions. Sample 4 and 5, respectively, with nominal composition $Cu_{0.9}CrS_2$ and $Cu_{0.8}CrS_2$, were made by the inter diffusion of binaries and annealing at 850$^0$C for 2 days.

The resistances of all the polycrystalline pellets were measured parallel to their planes and were found to increase on cooling to low temperatures. Sample (1 and 2) with the maximum flaky texture has substantially less resistivity at room temperature. A characteristic feature, viz. the plateau in their resistivity can be seen to develop around 150-200K. This feature was absent in phases with larger resistances which were prepared at lower temperatures, especially by using inter diffusion route. We believe that the plateau-like feature in the resistance on cooling is a characteristic feature of the phases obtained after extended period of sintering at high temperature that gives a high degree of flaky texture (see figure 1) and substantially smaller value of resistivity at room temperature. This behavior may be due to carrier doping and related to yet unknown nature of localization by impurities on cooling in their cation-rich compositions. The resistivity values at 300K for different samples are tabulated in Table 1.

A common feature for all the samples, including Cu-deficient ones, is the non-insulating like temperature dependence at low temperatures, as can be seen in figure 3. Although a significant increase in resistance on cooling towards magnetic transition is found, this increase is surely non-exponential. The resistance at $T_N$ becomes at most 10 to 20 times the room temperature value, and then increases much more slowly at low temperatures in the antiferromagnetic phase. The resistivity at $T_N$ exhibits hysteretic anomaly, as can be seen in the inset of panel "a" of figure 3. This confirms the first order nature of this transition which was also reported in the recent neutron scattering study[15]. They found a structural modification (Rhombohedral to Monoclinic) due to a strong magnetocrystalline coupling in this compound resulting in a first order nature of transition at $T_N$ [15]. A previous study on the isostructural-isoelectronic compound $AgCrS_2$ also reported a diffused first-order nature of antiferromagnetic-paramagnetic phase transition at 42K by heat capacity measurement[16].

In the lower panel of figure 3, we have shown the resistance measured in-plane of the crystal flakes of $CuCrS_2$. We find a metal-like dependence in all of them and large variations in the resistivity of different flakes from 5mΩ-cm to 100mΩ-cm around room temperature. The resistance shows small rise after it passes through a Kondo-like minimum below 100K. The transitional anomaly as a downward jump in resistance at magnetic transition is also seen in some of the flakes. Our results clearly show that the resistivity behavior of $CuCrS_2$ is quite complex and is unlike a doped-semiconductor as was suggested in earlier studies [9, 12].

***Thermopower:*** The most remarkable property of $CuCrS_2$ is the abnormally large value of Seebeck coefficients $S$ of our pellets and a crystal flake as shown in figure 4. The samples obtained after extended sintering at high temperatures (1, 2 and 3) have comparatively higher values of $S$ with the maximum 450μV/K for the quenched phase. The compounds prepared through inter-diffusion route $Cu_{0.9}CrS_2$ (no.4) and $Cu_{0.8}CrS_2$ (no.5) have comparatively smaller values, although $S$ remains significantly large at 200-300 μV/K around room temperature. The room temperature value of $S$ for different samples is presented in Table 1. The sign of $S$ is positive for all samples including the flakes. Moreover, $S$ is found to vary smoothly across the magnetic transition temperature $T_N$; it increases rapidly on heating above the magnetically ordered phase and saturates to a constant value in the paramagnetic phase above ~150K.

***Thermal conductivity:*** In figure 5, we have shown the thermal conductivity $\kappa$ of different pellets as a function of temperature by plotting it on log ($T$) scale between 15 K and 300K. The thermal conductivity of a sintered Alumina substrate, measured for the calibration purpose of our apparatus, is shown in the inset of figure 5. The value of $\kappa$ for different samples at room temperatures is quite different. This variation in $\kappa$ value may partly be due to varying contributions of the grain boundary scattering of the phonons due to their rather poor packing density (80% - 85% in our case) for sintered pellets. The $\rho$ and $S$ behavior of the same pellets have been displayed in figure 3 and 4. It is remarkable to find that among them the value of $\kappa$ is substantially less for the quenched phases (1 and 2), which have comparatively smaller resistivity and larger thermopower at room temperature. We note that the contribution to the thermal conductivity in these compounds is mostly due to the atomic lattice. The electronic contribution, as estimated using Lorenz number, accounts for less than 30% of $\kappa$ value at room temperature even in the case of most conducting pellets (1 and 2). Moreover, at low temperatures the electronic contribution becomes insignificant in all the samples. The further reduction of $\kappa$ value in the quenched phase may be the result of increased atomic disorder in the occupancy of Cu – atoms.

An unusual and remarkably large effect of magnetic transition on the thermal conductivity can be seen in all the pellets. We find that $\kappa$ shows a large dip as we approach magnetic transition $T_N$ from above as well as from below. We believe that the dip in $\kappa$ is caused by unusually large magnetoelastic coupling[15] that gives rise to the first order nature of the magnetic transition. As a result of the magnetoelastic coupling, short range magnetic correlations cause strong scattering of the phonons leading to a dip in $\kappa$ as the magnetic transition is approached from the paramagnetic side. Thermal conductivity $\kappa$ increases rapidly as soon as the magnetic order is set up. The strong reduction in $\kappa$ seems also to be correlated with the reduction in the electronic conduction (i.e., rise in resistivity; see figure 3) on cooling as we approach the magnetic transition. It is interesting to note that in this same temperature interval, the paramagnetic susceptibility also shows significant deviation from the Curie–Weiss behavior as can be seen from figure 2. These properties indicate the importance of magnetoelastic coupling on the scattering of electrons and phonons by the short range magnetic correlations which are built up much above $T_N$ in these layered compounds and cause simultaneous variations in the $\chi$, $\rho$, $S$ and $\kappa$ as observed.

## Discussion

The resistances of all pellets increase on cooling, in the same manner as found by Le Nagard in the case of single crystals[9]. Despite this increase, the resistivities at low temperatures remain quite small indicating the non-insulating character of our compounds. We believe this to be intrinsic property of these compounds. In this context, the observation of minimum in the resistivity of the flakes (see figure 3b) is quite significant, and indicates a Kondo-type magnetic scattering of conduction electrons from the paramagnetic centers. This mechanism can also explain unusually large Seebeck coefficient $S$ at high temperatures. In the case of Kondo model, the exchange interaction between localized paramagnetic center and the conduction electrons causes scattering which depends strongly on the energy of the conduction electrons, thus leading to anomalously large and temperature independent thermopower at high temperatures. Our results also show that the qualitative behavior of the electronic transport is not much affected by the change in carrier doping and the disorder in the Cu-occupancy in different metal rich compositions obtained after long sintering at high temperatures and quenched through order – disorder transition.

The temperature dependence of electrical conductivity together with the exceptionally high value of thermopower in our compounds is very similar to that found in the Kondo-insulators, for example $Ce_3Pt_3Sb_4$[17]. We therefore believe that a doped Kondo-insulator like phase is realized in the paramagnetic phase of $CuCrS_2$. In a Kondo-insulator conduction electrons suffer a small hybridization gap due to the proximity of the nearly localized 4*f*-energy-levels of rare-earth atom or, 3*d*-energy levels of Cr in our case. This causes a small excitation gap for the conduction of electrons, and in turn is also responsible for their inelastic scattering by the magnetic fluctuation of localized electrons. Strong energy dependence of this scattering gives high thermoelectric coefficient. However, in contrast to the non–magnetic properties of Ce-compounds, direct Cr-Cr magnetic interactions give rise to antiferromagnetic order in $CuCrS_2$ at low temperatures.

In Kondo-insulator, magnetic centers have unstable valency and show strong fluctuations of moments. From the observed Curie – Weiss (CW) dependence of magnetic susceptibility, nearly integral valency of Cr-atoms is deduced which suggests highly localized spin fluctuations of the Cr-atoms above 200K. At lower temperatures $\chi$ deviates significantly from CW behavior, as can be seen in figure 2. Moreover, the neutron diffraction study [8] gives about 20% reduced moment of Cr-atoms than what is expected from spin only 3/2 value. The reduction in magnetic moment of Cr-atom may imply an efficient mixing of 3*d*-electrons with the valence electrons (or holes) of chalcogen atoms as the temperature is reduced. This mixing mechanism may be the reason behind the complex temperature dependence of transport properties of $CuCrS_2$, as observed by us.

In contrast to the layered compound $CuCrS_2$, the related cubic thiospinels compounds $CuCr_2S_4$ ($Se_4$, $Te_4$) show good metallic properties. In the latter compounds, Cr-atoms are clearly in intermediate valence state, as was deduced from the Curie constant in their paramagnetic phase and also from the saturation magnetization in the ferromagnetic phase [5]. In addition, the behavior of $\rho$ and $S$ in the metallic thiospinels are quite typical of *d*-band conductors, similar to the transition metal compounds, such as ferromagnetic nickel [5]. The qualitative difference in the transport properties of the layered counterpart is basically due the anisotropic structure and the smaller extent of broadening of 3*d*-band by the hybridization with the valence band originating from the sulfur anions. This leads to strong localization effects due to disorder induced scattering of conduction electrons. We have also done a detailed investigation of the properties of corresponding layer-selenide compound $CuCrSe_2$, which is also Antiferromagnetic below 55K. Our measurements show that, compared to the sulfide, $CuCrSe_2$ shows good metallic conduction and smaller value of thermopower, most probably due to increased hybridization broadening of 3*d*-bands with the valence band of selenium in this compound[18].

## Conclusion

We have prepared the layered compound $CuCrS_2$ by different heat treatments and have reported for the first time the thermoelectric power and thermal conductivity properties. A common feature of the transport properties of different samples is non-metallic, non-insulating dependence of resistivity at low temperatures and an unusually large value of Seebeck coefficient around room temperature. We discuss this behavior in terms of doped Kondo-insulator like phase in the paramagnetic phase of $CuCrS_2$. We find an unusually small activation for conduction and a highly energy dependent scattering of conduction electrons from the magnetic fluctuations which lead to anomalously large thermopower in $CuCrS_2$. These properties combined with their poor thermal conductivity, which is further reduced by increased disorder in Cu–occupancy in the quenched phases, make them potential candidates for various thermoelectric applications. These compounds also have large magnetocrystalline coupling. Consequently, both the conduction electrons and the phonons are strongly scattered by the correlated spin fluctuations of the Cr-atoms on cooling towards antiferromagnetic transition temperature. As a result of these scatterings the electrical conductivity as well as thermal conductivity is found to decrease while approaching magnetic transition temperature. A thorough study of the electronic structure and the low energy excitations in this interesting class of Layered –Antiferromagnets would be quite rewarding.

## Acknowledgement:

G. C. Tewari and T.S. Tripathi acknowledge the Council of Scientific and Industrial Research (CSIR) India. We deeply acknowledge the help and valuable suggestions of Dr G. Jeffrey Snyder Materials Science, California Institute of Technology, California. We also acknowledge Dr. Alok Banerjee UGC-DAE Consortium for Scientific Research (CSR), Indore, India for the magnetic measurements at the 14T-facility.

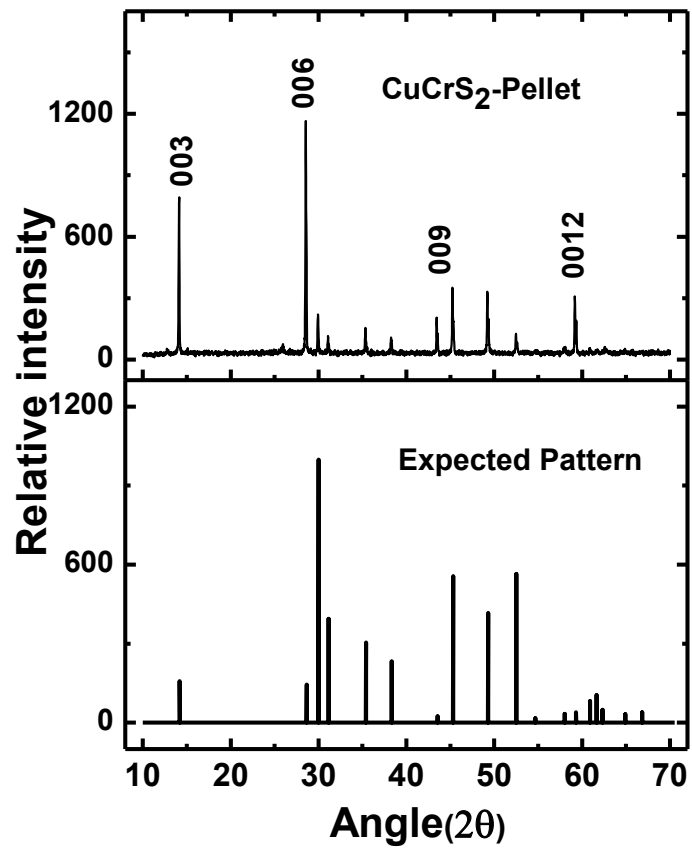

**Fig.1.** X-ray diffraction pattern from the surface of a polycrystalline pellet prepared by sintering above 850$^0$C over extended period. Compared to the powder pattern[9] it shows increased intensity of (00l) reflections due to platy-texture of the surface of the pellet.

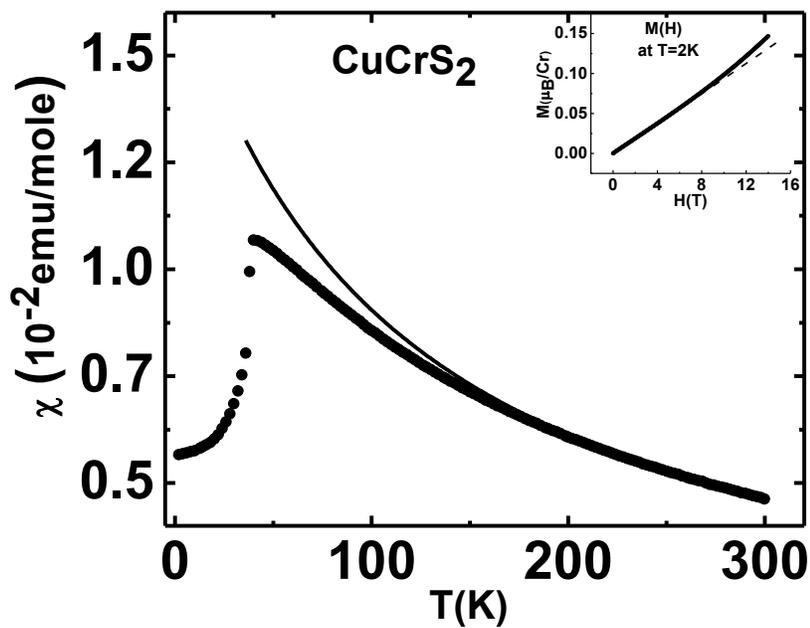

**Fig.2.** The magnetic susceptibility $\chi$ of CuCrS$_2$ showing departure from the Curie-Weiss dependence (continuous curve) below 200K. The upward deviation of magnetization *M* at 2K above 3Tesla field (shown in inset) indicates helicoidal order of Cr-spins in the antiferromagnet phase.

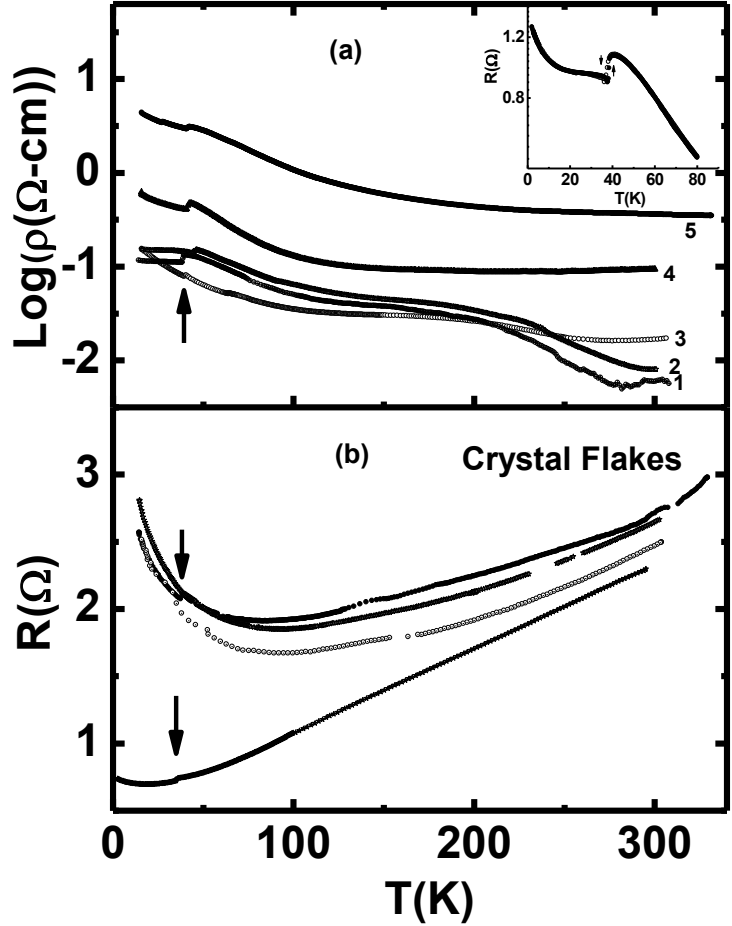

**Fig.3.** Panel (a), The resistivity dependence (plotted as Log($\rho$) versus $T$) of pellets of $CuCrS_2$ prepared by different thermal treatments. The extended period of sintering at higher temperature (1, 2 and 3) and quenching (1 and 2) gives lower resistivity compared to the phases prepared from the binary compounds including compounds $Cu_{0.9}CrS_2$ (4), $Cu_{0.8}CrS_2$ (5). Panel (b), Resistance of few crystal flakes showing minimum below 100K. Arrows mark magnetic transition $T_N$. The anomaly with the hysteresis at $T_N$ is shown for one of the pellets in the inset of panel (a).

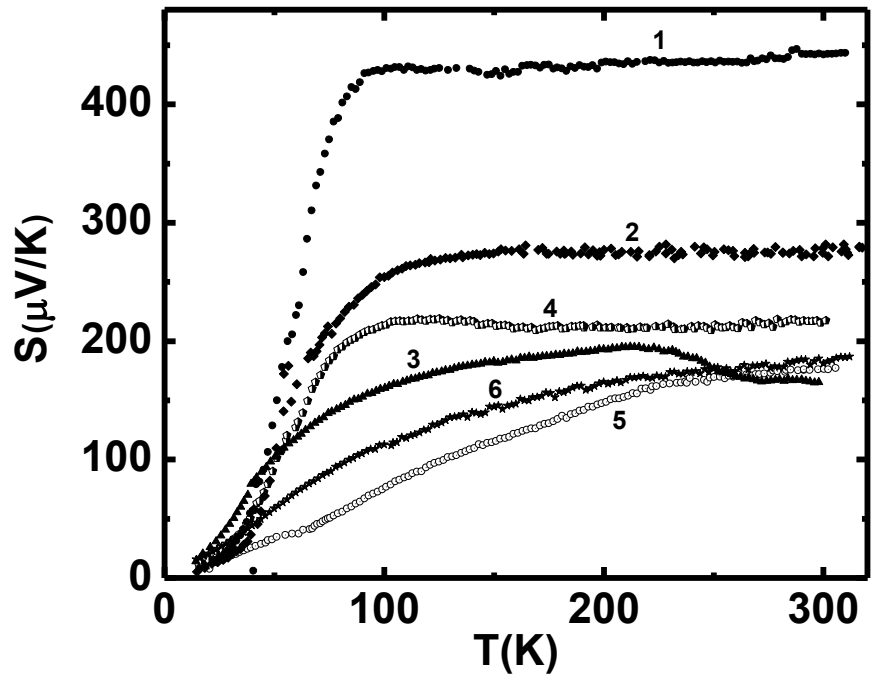

**Fig.4.** The positive Seebeck coefficients $S$ of different pellets and a crystal flake (6) showing large and constant values at high temperatures. The largest value of $S$ is found for pellets showing largest flaky texture. The temperature dependence of $S$ is similar for the phases with Cu-deficient composition $Cu_{0.9}CrS_2$ (4) and $Cu_{0.8}CrS_2$ (5).

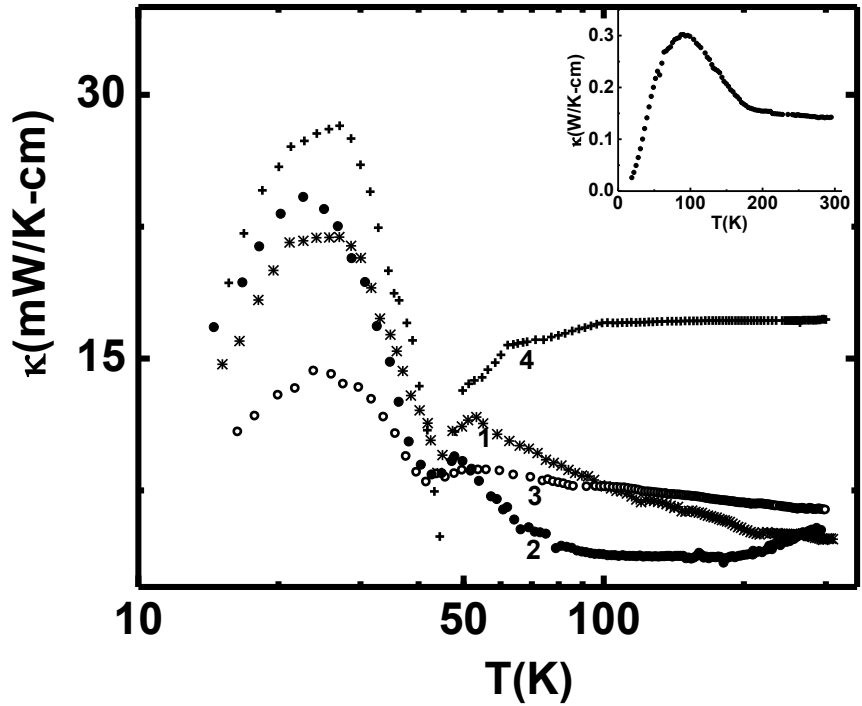

**Fig.5.** The thermal conductivity $\kappa$ of pellets plotted versus log($T$) for the quenched (1 and 2), slowly-cooled (3) and $Cu_{0.9}CrS_2$ (4). Strong dip in $\kappa$ is seen near $T_N$ in all of them. The $\kappa$ value in quenched phases (1 and 2) is reduced due to disorder in Cu-occupancy. The $\kappa$ measured on a thin sintered rectangular Alumina piece for the calibration of apparatus is presented in the inset.

Table1: Properties of polycrystalline pellets and flakes of $CuCrS_2$ at 300K.

| No. | Composition (nominal ratio) | Thermal Treatment[§1] | $\rho$ (m$\Omega$-cm) | $S$ ($\mu$V/K) | $\kappa$ (mW/(K-cm)) | $ZT$[§3] |
|---|---|---|---|---|---|---|
| 1 | $CuCrS_2$ ( Cu+Cr+2S ) | Extended sintering (5days) at $850^0$C + air quenched | 6 | 445[§2] | 4.8 | 2.0 |
| 2 | $CuCrS_2$ (Cu+Cr+2.1S) | Extended sintering (5days) at $950^0$C+ air quenched | 8 | 276[§2] | 5.4 | 0.53 |
| 3 | $CuCrS_2$ (Cu+Cr+2.2S) | Extended sintering (5days) at $850^0$C + slowly cooled | 15 | 170 | 6.5 | 0.1 |
| 4 | $Cu_{0.9}CrS_2$ ($0.9CuS+0.5Cr_2S_3$) | Sintered (2days) at $900^0$C +air quenched. | 90 | 217 | 17 | 0.01 |
| 5 | $Cu_{0.9}CrS_2$ ($0.8CuS+0.5Cr_2S_3$) | Sintered (2days) at $900^0$C +air quenched. | 350 | 176 | ------ | ------ |
| 6 | Crystal Flake | Excess S-vapor (thickness ~20-30$\mu$m). | 8 | 185 | ------- | ------ |

§1  packing density of the sintered pellets is about 80-85% of the theoretical value.
§2 Pellets with large flaky texture.
§3 $ZT=S^2T/(\rho\kappa)$